\newcommand{\CLASSINPUTinnersidemargin}{18mm}
\newcommand{\CLASSINPUToutersidemargin}{12mm}
\newcommand{\CLASSINPUTtoptextmargin}{20mm}
\newcommand{\CLASSINPUTbottomtextmargin}{25mm}
\documentclass[10pt,conference,a4paper]{IEEEtran}

\usepackage[english]{babel}
\usepackage[utf8x]{inputenc}

\renewcommand{\thesubsection}{\Alph{subsection}}
\renewcommand{\thesubsubsection}{\arabic{subsubsection}}
\addto\captionsspanish{\renewcommand\tablename{T\scshape{abla}}}
\addto\captionsspanish{\renewcommand\figurename{Fig.}}

\usepackage{times}

\usepackage{graphicx}
\usepackage{multirow}
\usepackage{xcolor}
\usepackage{booktabs}
\usepackage{amsmath}
\DeclareMathOperator*{\argmax}{argmax} 

\usepackage{amsmath}
\usepackage{times}
\usepackage{graphicx}
\usepackage{amsfonts}
\usepackage{amssymb}
\usepackage{cite}
\usepackage{subfigure}
\usepackage{url}
\usepackage{fancyhdr}

\DeclareGraphicsExtensions{.png,.eps,.ps,.pdf}

\begin{document}

\fancypagestyle{firstpage}
{
    \fancyhead[L]{Accepted at URSI 2020, Malaga}    
    \fancyhead[R]{2 - 4 September, 2020}
}

\title{Ensemble Methods and Input Alternatives for Acoustic Scene Classification Using Convolutional Neural Networks}


\author{
\authorblockN{Sergi Perez-Castanos$^{(1)}$, Javier Naranjo-Alcazar$^{(1,2)}$, Pedro Zuccarello$^{(1)}$, Maximo Cobos$^{(2)}$ and Francesc J. Ferri$^{(2)}$}
\authorblockA{[sergi.perez, javier.naranjo, pedro.zuccarello]@visualfy.com, [maximo.cobos, francesc.ferri]@uv.es}
\authorblockA{$^{(1)}$Visualfy, Benisan\'o, Val\`encia, Spain}
\authorblockA{$^{(2)}$Departament d'Inform\`atica, Universitat de Val\`encia, Burjassot, Spain}
}

\maketitle

\begin{abstract}
Acoustic scene classification (ASC) has been approached in the last years using deep learning techniques such as convolutional neural networks or recurrent neural networks. Many state-of-the-art solutions are based on image classification frameworks and, as such, a 2D representation of the audio signal is considered for training these networks. Finding the most suitable audio representation is still a research area of interest. In this paper, different log-Mel representations and combinations are analyzed. Experiments show that the best results are obtained using the harmonic and percussive components plus the difference between left and right stereo channels, $(L-R)$. On the other hand, it is a common strategy to ensemble different models in order to increase the final accuracy. Even though averaging different model predictions is a common choice, an exhaustive analysis of different ensemble techniques has not been presented in ASC problems. In this paper, geometric and arithmetic mean plus the Ordered Weighted Averaging (OWA) operator are studied as aggregation operators for the output of the different models of the ensemble. 
Finally, the work carried out in this paper is highly oriented towards real-time implementations. In this context, as the number of applications for audio classification on edge devices is increasing exponentially, we also analyze different network depths and efficient solutions for aggregating ensemble predictions.
\end{abstract}

\thispagestyle{firstpage}

\section{Introduction}\label{sec:intro}

\par Sounds carry a large amount of information about everyday environments. Therefore, developing methods to automatically extract this information has a huge potential in relevant applications, such as autonomous cars or home assistants. In \cite{han2016acoustic}, an audio scene is described as a collection of sound events on top of some ambient noise. Given a predefined set of tags where each describes a different audio scene (i.e. airport, public park, metro, etc.) and an audio clip coming from a particular audio scene, Audio Scene Classification (ASC) is the automatic assignment of one single tag to describe the content of the audio clip.
This problem has attracted the interest of the audio processing community in the last years, as evidenced by the first Task of the DCASE (Detection and Classification of
Acoustic Scenes and Events) 2019 Challenge \cite{Mesaros2018_DCASE}, which encouraged the participants to propose different solutions to tackle the ASC problem in a public tagged audio dataset.


\par Classic ASC systems were based on feature-engineering approaches, where the research effort was mainly aimed at developing meaningful features and using them to feed classical classifiers, such as GMMs or SVMs \cite{martin2016case}. Over the last years, Deep Neural Networks (DNNs) and, particularly, Convolutional Neural Networks (CNNs) have shown remarkable results in many different areas \cite{alexnet,bhandare2016applications}, thus being the most popular choice among researchers and application engineers. CNNs allow to solve both problems -feature extraction and classification- simultaneously within a single computational structure. 

Although several works for automatic audio classification have successfully proposed to feed CNNs with raw 1D audio signals \cite{lee2018samplecnn}, most state-of-the-art approaches use 2D time-frequency representations as a suitable input representation \cite{su2019environment, cakir2016domestic}. This also creates the need for setting up appropriate parameters (e.g. window type, size and/or overlap). Nonetheless, the advantage of 2D time-frequency representations is that they can be treated and processed with CNNs that have shown successful results with images. 


\section{Method}
\label{sec:method}

In this section, the general methodology followed in this work is described. First, a brief background on CNNs is provided, explaining the most common layers used in their design. Then, we explain the audio pre-processing steps considered for feeding the network, as well as its internal architecture. Finally, the ensemble techniques considered in this work are summarized. 

\subsection{Convolutional Neural Networks}\label{subsec:cnn}

\par The main feature of a CNN is the presence of convolutional layers that perform filtering by shifting a small window (receptive field) across the input signal, either 1D or 2D. These windows contain kernels that change their values during training according to a cost function. Activations computed by each kernel are known as features maps and they represent the output of the convolutional layer.

\par Other layers commonly used in CNNs are Batch Normalization and Dropout. These layers are interspersed between the convolutions to achieve a greater regularization and generalization of the network. Batch Normalization is an intermediate layer that normalizes kernel outputs during training. It is usually stacked between the convolutional and activation layers. Dropout tries to reduce overfitting during training by randomly disabling neurons at a specific rate.
Dropout layers are commonly stacked after pooling layers. These last layers subsample feature maps, thus reducing its dimension, usually, to the highest value of the receptive field (max pooling). It is a common practice to increase the number of filters in the convolutional forthcoming layers after a pooling layer. Thanks to these layers, CNNs are able to obtain higher level features adapted to the input data. 


\subsection{Audio preprocessing}\label{subsec:audio}

\par A combination of the time-frequency representations detailed in Table~\ref{tab:spect_features} has been used as input to the CNN (see Table~\ref{tab:results_dev}). All of them are based on the log-Mel spectrogram \cite{kong2018dcase} with 40~ms of analysis window, 50\% of overlap between consecutive windows, Hamming asymmetric windowing, FFT of 2048 points and a 64-band normalized Mel-scaled filter bank. Each row, corresponding to a particular frequency band, of this Mel-spectrogram matrix is then normalized according to its mean and standard deviation. 
Another type of input representation considered in this work are the extracted harmonic and percussive components using the Harmonic-Percussive Source Separation (HPSS) algorithm \cite{Fitzg:2010:harmo}, using spectrograms with the same aforementioned parameters.


Considering that the audio clips are 10~s long, the size of the final log Mel feature matrix is $64 \times 500$ for all the cases. The audio preprocessing has been developed using the \mbox{LibROSA} library for Python. 

\begin{table*}[t!]
\caption{Input representations used in this work. Several combinations have also been tested as input to the CNN (see Table~\ref{tab:results_dev} and Sec.~\ref{sec:results}).}
\centering
\begin{tabular}{c|l|l}
 & & Description\\
\toprule
M & Mono       & Log Mel spectrogram (computed as detailed in Sect.~\ref{subsec:audio}) of the arithmetic mean of the left and right audio channels.\\
L & Left       & Log Mel spectrogram (computed as detailed in Sect.~\ref{subsec:audio}) of the left audio channel.\\
R & Right      & Log Mel spectrogram (computed as detailed in Sect.~\ref{subsec:audio}) of the right audio channel.\\
D & Difference & Log Mel spectrogram (computed as detailed in Sect.~\ref{subsec:audio}) of the difference of the left and right audio channels $(L-R)$.\\
H & Harmonic   & Harmonic Log Mel matrix (computed as detailed in Sect.~\ref{subsec:audio}) using the Mono signal as input.\\
P & Percussive & Percussive Log Mel matrix (computed as detailed in Sect.~\ref{subsec:audio}) using the Mono signal as input.
\end{tabular}
\label{tab:spect_features}
\end{table*}

\begin{table*}[t!]
\caption{Network accuracy (\%) for the development set (Dev set). The accuracy was calculated using the first evaluation setup of 4185 samples. The models labeled with an (*) were used for ensembles (see Table~\ref{tab:ev_tab}). PN indicates the number of parameters in the network.}
\centering
\begin{tabular}{@{}ccccc@{}}
\toprule
\multicolumn{2}{c}{Audio preprocessing} & \multicolumn{3}{c}{Networks}  \\ \cmidrule(lr){1-2} \cmidrule(lr){3-5}
Audio representation & Channels & \textit{\textbf{Vfy-3L16}} & \textit{\textbf{Vfy-3L32}} & \textit{\textbf{Vfy-3L64}} \\ \midrule \midrule
&  & Dev set & Dev set & Dev set \\ 
\midrule
\multirow{9}{*}{Log Mel spectrogram} & Mono (M) & \textit{70.47}   & 70.07  & 70.49  \\ \cmidrule{2-5} 
& Left + Right + Difference (LRD) & 72.69  & \textit{*73.76}   & 73.41  \\ \cmidrule{2-5} 
& Harmonic + Percussive (HP) & 71.23  & \textit{71.85}  & 72.04 \\  
\cmidrule{2-5} 
 & Harmonic + Percussive + Mono (HPM)  & 69.37  & 70.99   & 71.59  \\  
\cmidrule{2-5} 
  & Harmonic + Percussive + Difference (HPD)  & 72.64 & \textit{*\textbf{75.75}}   & 75.44 \\ 
\cmidrule{2-5} 
  & Harmonic + Percussive + Left + Right (HPLR)  & 71.57 & \textit{*72.76}    & 73.19 \\   
 \midrule
  &  & \textbf{PN (3D): 176,926} & \textbf{PN (3D): 495,150} & \textbf{PN (3D): 1,560,142} \\
  \bottomrule
\end{tabular}
\label{tab:results_dev}
\end{table*}

\subsection{Network Architecture}\label{subsec:network}

\par The network proposed in this work is inspired in the architecture of the VGG \cite{simonyan2014very}, since this last one has shown successful results in ASC \cite{su2019environment,zhang2019acoustic}. 
The convolutional layers were configured with small $(3 \times 3)$ receptive fields. After each convolutional layer, Batch Normalization and Exponential Linear Units (ELUs) activation layers \cite{clevert2015fast} were stacked. Two consecutive convolutional layers, including their respective Batch Normalization and activation layers, plus a max pooling and dropout layer correspond here to a \emph{convolutional block}. The final network (see Table~\ref{tab:net_arch}) is composed of three convolutional blocks plus two fully-connected layers acting as classifiers.

\par Three different values for the number of filters for the first \textit{convolutional block} have been implemented and tested: 16, 32, and 64 (see Table~\ref{tab:results_dev}). The developed network is intended to be used in a real-time embedded system, therefore a compromise has been achieved between the number of parameters and the final classification accuracy.

\begin{table}[]
\caption{Network architecture proposed for this challenge. The name indicates the number of convolutional blocks in the network and the number of filters of the first convolutional block.}
\centering
\begin{tabular}{c}
\toprule
\textbf{Visualfy Network Architecture - Vfy-3L$X$}             \\
\toprule
{[}conv (3x3, \#$X$), batch normalization, ELU(1.0){]} x2  \\
\midrule
MaxPooling(2,10)                                        \\
\midrule
Dropout(0.3)                                            \\
\midrule
{[}conv (3x3, \#2$X$), batch normalization, ELU(1.0){]} x2 \\
\midrule
MaxPooling(2,5)                                         \\
\midrule
Dropout(0.3)                                            \\
\midrule
{[}conv (3x3, \#4$X$), batch normalization, ELU(1.0){]} x2    \\
\midrule
MaxPooling(2,5)                                         \\
\midrule
Dropout(0.3)                                            \\
\midrule
Flatten                                                 \\
\midrule
{[}Dense(100), batch normalization, ELU(1.0){]}          \\
\midrule
Dropout(0.4)                                            \\
\midrule
{[}Dense(10), batch normalization, softmax{]}  \\
\bottomrule
\end{tabular}
\label{tab:net_arch}
\end{table}

\subsection{Model Ensemble Techniques}\label{subsec:ensemble}

\par Combining predictions from different classifiers has become a popular technique to increase the accuracy of the system. These combinations, known as \emph{ensemble models} \cite{silla2007automatic}, can be approached in two different ways: in the first one, known as \emph{stacking}, a second classifier is trained so that it learns from the intermediate representations of a first set of classifiers. For example, in  \cite{sakashita2018acoustic}, Sakashita \textit{et al.} trained a Random Forest using the representations of 9 different models. A second approach to model ensembles is to calculate the final prediction using the predictions of several classifiers combined by a specific fusion technique. So far, the most common choice is the arithmetic mean. In this case the final prediction, $\widehat{c}$, would be: 

\begin{equation}
\widehat{c} = \argmax_{c \in \mathcal{C}} \left(\frac{1}{M}\sum^M_{m=1} P_m(c|\bar{x})\right),
\end{equation}
where $\mathcal{C}$ is the number of classes, $M$ is the number of models to be ensembled, and $P_m(c|\bar{x})$ is the prediction likelihood for class $c$ of model $m$ given a specific chunk $\bar{x}$ of audio.

Another option would be to apply \textit{and-like} operators like the geometric mean \cite{silla2007automatic}. In this case, $\widehat{c}$, is calculated as:
\begin{equation}
\widehat{c} = \argmax_{c \in \mathcal{C}} \sqrt[M]{\prod^M_{m=1} P_m(c|\bar{x})}.
\end{equation}

\par The final approach studied in this paper is the ordered weighting averaging (OWA) operator \cite{leon2007applying}. In this case the probabilities are sorted from highest to lowest. Then, a weighted sum with a predefined set of weights is performed. In this way, the $i$-th weighting coefficient is not assigned to the output of a particular model but instead is assigned to the $i$-th output value after the values have been sorted. Therefore, the OWA operator is not linear. Depending on the chosen set of weights the operator can be \textit{or-like} or \textit{and-like}. This is, more weight can be assigned to the largest probabilities (\textit{or-like}) or to the lowest (\textit{and-like}). See \cite{leon2007applying} for an in-depth discussion about the usage of OWA operator in the context of image retrieval systems. In this last case, the final prediction would be:
\begin{equation}
\widehat{c} = \argmax_{c \in \mathcal{C}} \left(\sum^M_{i=1} w_i P_i(c|\bar{x})\right).
\end{equation}
where $w_i$ is the weighting coefficient associated to the $i$-th largest output probability and the weights fulfill the constrain $\sum_{i=1}^M w_i=1$. In the context of this work the number of aggregated models is $M=3$. The OWA weighting vector is $W=[0.1,\, 0.15,\, 0.75]$. It is clear that $W$ has been configured as an \textit{and-like} operator giving more relevance to the lowest probability value.

\begin{table*}[t]
\caption{Final results (\%) of DCASE 2019 Task 1a challenge.}
\centering
\begin{tabular}{cccccc}
\toprule
\textbf{Model name}               & \textbf{Ensemble method} & \textbf{Ev. acc} & \textbf{Dev. acc} & \textbf{Ev. acc on seen cities} & \textbf{Ev. acc on unseen cities} \\

\midrule \midrule

Naranjo-Alcazar\_VfyAI\_task1a\_2 & Geom. mean  & \textbf{74.2} & \textbf{77.1} & \textbf{75.9} & \textbf{65.8} \\ \midrule 
Naranjo-Alcazar\_VfyAI\_task1a\_4 & OWA & 74.1 & 76.9  & 75.8 & 65.7 \\ \midrule
Naranjo-Alcazar\_VfyAI\_task1a\_1 & Arith. mean  & 74.1 & 76.8 & 75.8  & 65.7 \\ \bottomrule
\end{tabular}
\label{tab:ev_tab}
\end{table*}

\section{Results}\label{sec:results}

\subsection{Experimental details}\label{subsec:details}

The optimizer used was Adam \cite{kingma2014adam} configured with $\beta_1 = 0.9$, $\beta_2 = 0.999$, $\epsilon = 10^{-8}$, $decay=0.0$ and $amsgrad = \mathrm{True}$  . The models were trained with a maximum of 2000 epochs. Batch size was set to 32. The learning rate started with a value of $0.001$ decreasing with a factor of 0.5 in case of no improvement in the validation accuracy after 50 epochs. If validation accuracy does not improve after 100 epochs, then training is early stopped. Keras with Tensorflow backend was used to implement the models. All the experiments consider the DCASE 2019 Task 1 dataset. The dataset is composed by audio recordings from 12 cities. Only 10 of them (seen cities) are used in development (Dev) stage. These 10 cities and the other 2 (unseen cities) are taken into account in evaluation (Ev) stage.

\begin{table}[t]
\caption{Class accuracies (\%) on development stage.}
\centering
\begin{tabular}{ccccc}
\toprule
Class   & \multicolumn{4}{c}{Network}   \\
\midrule
& \multicolumn{3}{c}{E\_LRD\_HPD\_HPLR}   & Baseline \\ \cmidrule{2-5}
& Sum   & Prod          & OWA        &     \\ \cmidrule(lr){2-4} 
Airport   & \textbf{79.8} & 78.9  & 78.9          & 48.4     \\ 
\midrule
Bus       & 83.4     & \textbf{84.3} & 83.4     & 62.3     \\
\midrule
Metro    & 72.3     & \textbf{73.0} & \textbf{73.0} & 65.1     \\
\midrule
Metro station     & \textbf{76.1} & 74.9 & 75.6    & 54.5     \\
\midrule
Park    & 88.9    & \textbf{89.4}  & 89.1     & 83.1     \\
\midrule
Public square     & 61.0   & 61.0 & \textbf{61.2} & 40.7     \\
\midrule
Shopping mall     & 72.3   & \textbf{73.9} & 73.2   & 59.4     \\
\midrule
Street pedestrian & 78.1  & \textbf{78.6} & 78.3  & 60.9     \\
\midrule
Street traffic    & 90.3  & \textbf{90.5}    & 90.3  & 86.7     \\
\midrule
Tram              & \textbf{66.3}     & 66.1 & \textbf{66.3} & 64.0     \\
\toprule
Average           & 76.84    & \textbf{77.06} & 76.94         & 62.5 \\
\bottomrule
\end{tabular}
\label{tab:class_dev}
\end{table}

\subsection{Results on development set}\label{subsec:res_data}

Table~\ref{tab:results_dev} shows the results obtained for the development dataset when using the three networks detailed in Table~\ref{tab:net_arch} combined with the different inputs explained in Table~\ref{tab:spect_features}. Table~\ref{tab:results_dev} shows that when the network is fed with one channel input (M), the shallowest network shows the same results as the deepest. On the other hand, when the input is fed with more than one channel, deeper networks improve the accuracy, with different improvements depending on the selected audio representation. The most suitable representation is HPD. As far as this group is aware, this combination has not been proposed yet \cite{sakashita2018acoustic, han2017convolutional}. Ensemble accuracies per class are shown in Table~\ref{tab:class_dev}. 

\par On the other hand, when the ensembles detailed in Sect.~\ref{subsec:ensemble} are applied, the accuracy is also improved. The three models showing the highest accuracy in Table~\ref{tab:results_dev} have been used for this purpose. It is worth mentioning that although \textit{Vfy-3L64} shows, in some cases, better accuracy on the development set, this improvement is not correlated with the \emph{Public Leaderboard} set. An interpretation for this could be that the network is more prone to overfitting due to the highest number of parameters.

\subsection{Results on evaluation set}\label{subsec:res_ev}

Table \ref{tab:ev_tab} shows the final results of our systems on the evaluation and development stages. The ensemble using geometric mean shows the best performance. It can be observed that the cities whose audios have not been used for training show lowest accuracy values than the other ones, even though they both exhibit the same rank.

\section{Conclusions and future work}\label{sec:conclusion}

In order to embed an ASC classifier into an edge system, the depth of the network becomes a crucial factor. Real-time devices usually work under sharp constraints concerning the classification time. In this paper, a study of three different CNNs have been carried out. As expected, deeper networks show larger classification times, nevertheless, they do not always present better accuracy values. In the present study no substantial difference has been found, in terms of accuracy, between a network having 0.5M parameters and another one with 1.5M. Most CNN architectures proposed in the literature are very similar: $3\times3$, $5\times5$ or $7\times7$ kernel sizes with a selected number of filters, an structure that is repeated sequentially multiplying by 2 the number of filters with interleaved dropouts or max-pooling layers. Thus, the selection of the number of filters in the first layer may be an essential aspect that must be taken into account for real-time applications. Moreover, it must be emphasized that no data augmentation has been performed. Instead, this work has been centered on a proper analysis of different input alternatives and ensembles techniques (arithmetic mean, geometric mean and OWA). The results clearly show that a simple method, like the arithmetic mean of the output class likelihoods, outperforms the accuracy of a single model. The highest scores have been obtained with the geometric mean and the OWA considering an AND-like weighting vector. Given that the geometric mean and OWA are both AND-like operators, it may be concluded that prediction of the ensemble is high only when all the predictions of the different models are high. A more in-depth study of different OWA weights is left as future work. Finally, regarding the input audio representation, the use of separated harmonic and percussive components showed the best accuracy in the three networks, which reinforces the idea that selecting a proper representation can be more important than implementing computationally expensive networks.

\section*{Acknowledgment}

This project has received funding from the European Union’s Horizon 2020 research and innovation program under grant agreement No 779158. The participation of Javier Naranjo-Alcazar and Dr. Pedro Zuccarello in this work is partially supported by Torres Quevedo fellowships DIN2018-009982 and PTQ-17-09106 respectively from the Spanish Ministry of Science, Innovation and Universities. The participation of Dr. Maximo Cobos and Dr. Francesc J. Ferri was supported by FEDER and the Spanish Ministry of Science, Innovation and Universities under Grant RTI2018-097045-B-C21.

\bibliographystyle{IEEEtran}
\bibliography{refs}

\end{document}